\documentclass[twocolumn,showpacs,amsmath,amssymb,aps,prl,superscriptaddress]{revtex4}
\usepackage{graphicx}
\newcommand {\be}{\begin{equation}}
\newcommand {\ee}{\end{equation}}
\newcommand {\ben}{\begin{displaymath}}
\newcommand {\een}{\end{displaymath}}
\newcommand {\bea}{\begin{eqnarray}}
\newcommand {\eea}{\end{eqnarray}}
\newcommand {\nn}{\nonumber}

\begin{document}
\title{$d$-Mott phases in one and two dimensions}
\author{A. L\"auchli}
\affiliation{Laboratoire de Physique Th\'eorique, CNRS-UMR 5152, 
  Univ. Paul Sabatier, F-31062 Toulouse, France}
\affiliation{Theoretische Physik, ETH-H\"onggerberg, CH-8093 Z\"urich, Switzerland}
\author{C. Honerkamp}
\affiliation{Theoretische Physik, ETH-H\"onggerberg, CH-8093 Z\"urich, Switzerland}
\affiliation{Department of Physics, Massachusetts Institute of Technology, Cambridge MA 02139, USA}
\author{T.M. Rice}
\affiliation{Theoretische Physik, ETH-H\"onggerberg, CH-8093 Z\"urich, Switzerland}
\date{\today}
\begin{abstract}
  We use exact diagonalization to determine the spectrum of reduced Hamiltonians based 
  on renormalization group flows to strong coupling. For the half-filled 
  two-leg Hubbard ladder we reproduce the known insulating $d$-Mott groundstate with
  spin and charge gaps. For the saddle point regions of the two-dimensional Hubbard
  model near half-filling we find a crossover to a similar strong coupling state,
  which truncates the Fermi surface near the saddle points. At lower scales $d$-wave
  superconductivity appears on the remaining Fermi surface.
  %%  This suggests a partially insulating state where the 
  %%  Fermi surface near the saddle points becomes truncated and replaced with a $d$-wave 
  %%  paired Mott state.
\end{abstract}
\maketitle

%%{\em Introduction:}
In recent years the application of renormalization group (RG) methods to
interacting fermions in low dimensions has made great progress. In one
dimension  (1D) where the Fermi surface (FS) is a discrete
set of points, the method has a long history going back to the
pioneering work of Menyh\'ard and S\'olyom \cite{solyom}. In two
dimensions (2D) the FS is a continuous curve and the discrete
set of coupling constants is replaced by a 4-point vertex
which is a function of three independent momenta. Zanchi and Schulz \cite{zanchi} 
developed a numerical method to evaluate the functional renormalization 
group flow of the scattering vertex based on a break-up of the FS 
into $N$ patches. The resulting coupled differential equations can
be integrated numerically to obtain the RG flow. This method has been
applied to the Hubbard model on a square lattice with nearest-neighbor
(n.n.) and next-nearest neighbor (n.n.n.) hopping using a Wilsonian
scheme which successively integrates out high energy modes down to
an infrared cutoff \cite{zanchi,halboth,honerkamp}. The RG method has
the advantage that it allows an unbiased treatment of competing instabilities 
including even Stoner ferromagnetism, when the recent reformulation\cite{tflow}, 
with the temperature as the flow parameter, is used.

A serious limitation to RG schemes occurs when the scattering vertices flow to 
strong coupling as the RG scale is lowered. Although the one-loop flow may still 
be a good approximation even when the interaction strength has become comparable
to the kinetic energy, self energy corrections \cite{honerkamp} typically become 
important shortly after that and  higher order terms or non-perturbative effects 
might not be negligible any longer.  Thus these RG flows cannot be extrapolated to 
scales below the divergence. Generally those are the most interesting
cases as divergent RG flows signal a breakdown of the perturbative normal
state. In the special case of 1D, bosonization allows 
a complete analysis of the strong coupling regime. 
In higher dimensions, in cases where a single instability
dominates the RG flow, a mean field approximation
can be introduced to describe the strong coupling phase. 
%% possibly ommited
This can be corrected for fluctuations around the mean field to obtain a
satisfactory description of the ordered phase. 
Yet in some especially interesting cases, there can be several mutually
reinforcing instabilities so that it is not clear how to formulate a
mean field approximation. 

In this Letter we propose a new and more general approach to determine the
strong coupling phase that emerges on scales below divergent RG
flows. The starting point is to recognize that a mean field
approximation based on factorization of the interaction term in the
Hamiltonian, can also be expressed as an exact solution of a reduced
Hamiltonian e.g. BCS theory was formulated as an exact solution for a
reduced Hamiltonian in which only terms describing the
scattering of Cooper pairs with zero total momentum were kept \cite{BCS}.
Note these are the most divergent terms in the RG flow of the 4-point vertex. 
%% possibly ommited
Thus we can regard the mean field approximation as contained in the 
solution of a reduced Hamiltonian whose form is derived from the diverging RG flow.
In those cases mentioned above the reduced Hamiltonian is more complex and 
cannot be simply solved analytically. However it is amenable to numerical
diagonalization and this is the approach we take here. 
In particular in the study of a n.n.n. Hubbard model on a square lattice as a prototypical
model for underdoped cuprates, Furukawa et al.\cite{furukawa}, and Honerkamp et al.\cite{honerkamp},
found a {\em mutual reinforcement} of antiferromagnetic (AF) and $d$-wave pairing
tendencies when the Fermi energy was close to the van Hove singularities at the saddle points (SP) at 
$(\pm\pi,0),~(0,\pm\pi)$. They pointed out that the same mechanism is at work in the RG flows in the 
{\em half-filled} 2-leg Hubbard ladder (2LHL). At the heart of the mutual reinforcement is the
geometrical fact that in these cases a significant fraction of particle pairs that experience
strong Umklapp scattering in the $(\pi,\pi)$ channel, e.g. two particles close to the SP $(\pi,0)$,
have small total momentum and thus couple into the Cooper channel as well. Since both the AF 
$(\pi,\pi)$-scattering and the large momentum transfer pair scattering in the 
$d_{x^2-y^2}$-channel are repulsive, a mutual reinforcement occurs. In the 
half-filled 2LHL, this works down to arbitrarily low scales. In 2D, the overlap between the channels
decreases for low scales such that a finite interaction strength is needed for this mechanism.
The strong coupling behavior of the ladder has been fully analyzed using Density Matrix
Renormalization Group (DMRG)\cite{noack} and by bosonization by Lin et al.\cite{fisher} who 
named the groundstate a $d$-Mott phase. It shows the unusual
behavior of enhanced correlations simultaneously in the AF, $d$-wave pairing and $d$-density wave 
(or orbital AF) channels and is intimately related to resonating-valence bond (dRVB) states at large
$U$. Note all correlations remain strictly short range due to the presence of both spin and charge 
gaps. Clearly this is a case of a nonperturbative groundstate which cannot be even qualitatively 
described by a mean field approximation based on long range order.

We analyze first the case of the 2LHL which is a good test case since the groundstate is highly 
nontrivial and turn then to the n.n.n. Hubbard model on a square lattice.
The limitation of Lanczos algorithms to a relatively small number of $\vec{k}$-points means that in
2D a representation of the whole FS is impossible. However in the RG flows of the 2D n.n.n. Hubbard
model, the dominant divergent scattering processes appear in the vicinity
of the SPs. This allows us to restrict our reduced model, derived from the RG flow, to processes 
connecting a discrete set of $\vec{k}$ points located near the two SPs.
If this discrete set of $\vec{k}$-points is restricted to just the two
SPs, then the RG flow reduces to a small number of coupled
differential equations. Two different fixed points were found by
Dzyaloshinskii\cite{dzyaloshinskii} and by Lederer et al.\cite{lederer}. 
The latter found a divergent flow to strong coupling when the next to 
leading order particle-hole processes are included. 
The 2D RG flows have a similar form for the
dominant processes connecting the SP regions. Our
numerical results for this case reveal both spin and charge gaps at the saddle
points and show enhanced correlations in both AF and $d$-wave pairing
channels. They confirm the similarity to the $d$-Mott phase of the 2LHL, in line with the 
earlier interpretations.

\begin{figure}
  \centerline{\includegraphics[width=0.6\linewidth]{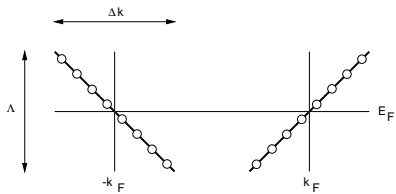}}   
\vspace{-1mm}
  \caption{
    Discretization of the $\Lambda$-shell around two Fermi points.
    Small circles denote individual $\vec{k}$-points. A set of $\vec{k}$-points
    belonging to the same branch is called a {\em patch}.
    \label{fig:DiscretizedShell}}
\end{figure}

%{\em The method ---}
Let us briefly describe the numerical approach. An extended discussion and results for the
2LHL and other 1D systems will be published elsewhere \cite{amlOnMethod}.
Our scheme is based on the numerical analysis of the reduced Hamiltonian $H_\Lambda$
obtained from the RG flow close to the critical scale $\Lambda_c$.
$H_\Lambda$, acting on states in a $\Lambda$-shell around the Fermi points or surface, is 
discretized using a finite number of $\vec{k}$-points (illustrated in Fig. \ref{fig:DiscretizedShell}
for two Fermi points). Generically it has the form
\begin{eqnarray*}
  \label{eqn:Hkspace}
  H_\Lambda &=&\sum_{\vec{k},\sigma}\varepsilon(\vec{k})\
  c_{\vec{k},\sigma}^\dagger c_{\vec{k},\sigma} \nn \\
  && + 
  \frac{ \lambda }{2 \Omega}
  \sum_{\vec{k}_1,\vec{k}_2,\vec{k}_3 \atop \sigma,\sigma' } 
  V_\Lambda(\vec{k}_1,\vec{k}_2,\vec{k}_3) %%\nn\\
  \ c_{\vec{k}_3,\sigma}^\dagger
  c_{\vec{k}_4,\sigma'}^\dagger c_{\vec{k}_2,\sigma'} c_{\vec{k}_1,\sigma}\nn
\end{eqnarray*}
where $\varepsilon(\vec{k})$ denotes the kinetic energy, $\lambda$ is a
global coupling constant, $\Omega$ the total volume, $V_\Lambda(\vec{k}_1,\vec{k}_2,\vec{k}_3)$ 
the discretized coupling function and $\vec{k}_4=\vec{k}_1+\vec{k}_2-\vec{k}_3$ (modulo umklapp) 
due to momentum conservation. The momentum dependence of the interaction function 
$V_\Lambda(\vec{k}_1,\vec{k}_2,\vec{k}_3)$ is determined by the RG couplings at scale $\Lambda$:
\[
  V_\Lambda(\vec{k}_1,\vec{k}_2,\vec{k}_3)=
  \tilde{g}[\Lambda] \left(
    \mbox{Patch}(\vec{k}_1),
    \mbox{Patch}(\vec{k}_2),
    \mbox{Patch}(\vec{k}_3)\right)\, .
\]
Here $\tilde{g}[\Lambda]$ denotes the ratios of the diverging couplings and $\mbox{Patch}(\vec{k})$ is the
patch index of $\vec{k}$-point $\vec{k}$. 
Note that in the reduced BCS Hamiltonian, only interactions $V_\Lambda(\vec{k}_1,\vec{k}_2,\vec{k}_3)$
with $\vec{k}+\vec{k}_2=0$ are kept. 
In contrast to this, here in order to describe the short-range $d$-Mott state, a whole set of 
scattering processes with nonzero $\vec{k}+\vec{k}_2 \approx 0$ must be kept in $H_{\Lambda}$.
An analogous width for $V_\Lambda(\vec{k}_1,\vec{k}_2,\vec{k}_3)$ for momentum transfers
$\approx (\pi,\pi)$ is necessary.
To incorporate this essential feature we assume that all scattering amplitudes with the same
patch indices take the same value. 
The global coupling constant $\lambda$ is in principle determined by
the the initial conditions and the final scale $\Lambda$. Here however we use it as a parameter
in order to investigate the interplay between kinetic energy and scattering processes and to
analyze the effects of the finite number of $\vec{k}$-points.

In the next step the discretized Hamiltonian is diagonalized using a Lanczos algorithm.
Due to momentum conservation (modulo umklapp scattering) a number $N_k\lesssim 20$ of orbitals 
can be treated. In this way energy gaps, static and dynamical 
correlation functions are easily accessible, similar to standard real space diagonalization
algorithms. As a future development the Lanczos scheme might be replaced by a DMRG algorithm,
thus possibly allowing more $\vec{k}$-points for a higher resolution on the FS.

%{\em The half-filled two-leg ladder ---}
We illustrate the method with an application to the two leg Hubbard ladder at half 
filling. We choose $t_{\perp}<2t_{\parallel}$ in order to have 4 Fermi points. The
RG equations have been derived and analyzed in \cite{fisher}. In particular for 
general repulsive initial interactions the couplings flow towards a fixed ray, i.e. a 
reduced Hamiltonian with fixed ratios among the diverging scattering vertices.

\begin{figure}
\vspace{6mm}
  \begin{center}
    \includegraphics*[width=0.85\linewidth]{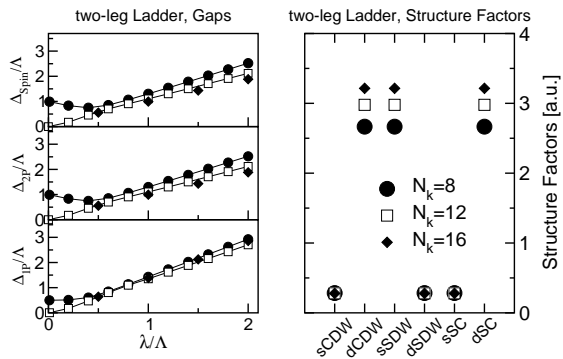}
  \end{center}   
\vspace{-3mm}
  \caption{
    Left panel: Spin, two particle and single particle gaps in the $d$-Mott
    phase of the two-leg Hubbard ladder as a function of the
    interaction parameter $\lambda$.
    Right panel: Order parameter susceptibilities for various kinds of
    order in the $d$-Mott phase of the two-leg Hubbard ladder.
    No evidence for long range order is found. However the
    staggered spin, the $d$-wave pairing and the staggered flux
    correlations are equally enhanced with respect to the 
    noninteracting ground state.
    \label{fig:DMottLadder} }
\end{figure}

We characterize the groundstate of the half-filled 2LHL first by
calculating different gaps: the spin gap: $\Delta_{\mbox{\tiny Spin}}=E_0(N_e,S^z=1)-E_0(N_e,S^z=0)$,
the single particle gap: $\Delta_{\mbox{\tiny 1P}}=[E_0(N_e-1,1/2)+E_0(N_e+1,1/2)]/2-E_0(N_e,0)$
and the two particle gap: $\Delta_{\mbox{\tiny 2P}}=[E_0(N_e-2,0)+E_0(N_e+2,0)]/2-E_0(N_e,0)$.
Results are shown in the left panel of Fig. \ref{fig:DMottLadder} for different system sizes and choices
of $\lambda$. One nicely sees that all gaps remain finite, even for $N_k\rightarrow \infty$. Interestingly the
spin gap and the two particle gap are exactly degenerate, while the single particle gap is slightly
larger. This can be understood as an approximate realization of the $SO(8)$ symmetry of the fixed ray,
discovered in Ref. \cite{fisher}. In our discretized scheme only a $SO(5)$ symmetry seems to be manifest.

%%%%%%%%%%%%%%%%%%%%%%%%%%%%%%%%%%%%%%%%%
%% Figure
%%%%%%%%%%%%%%%%%%%%%%%%%%%%%%%%%%%%%%%%%
%%%%%%%%%%%%%%%%%%%%%%%%%%%%%%%%%%%%%%%%%
%% Figure
%%%%%%%%%%%%%%%%%%%%%%%%%%%%%%%%%%%%%%%%%
Now we consider the correlation functions. We investigate particle-hole and particle-particle response
with both $s$-wave and $d$-wave formfactors. The $\vec{Q}=(\pi,\pi)$ charge density wave structure 
factors are defined as follows:
$
\mathcal{S}_{CDW} = \frac{1}{N_k} 
\left|
\sum_{\vec{k},\sigma} f(\vec{k}) c^{\dagger}_{\vec{k}+\vec{Q},\sigma} c_{\vec{k},\sigma} | \Psi\rangle 
\right| ^2,
$
where $f(\vec{k})= 1$ for the s-wave channel and $f(\vec{k})= \exp(i k_y)$ for the $d$-wave channel.
The structure factors for the spin density wave and the pairing correlations are defined similarly.
The results for the case $\lambda/\Lambda=2$ are shown in the right panel of 
Fig. \ref{fig:DMottLadder}, normalized to the respective expectation value in the noninteracting
$N_{\vec{k}}=8$ groundstate. The response for the standard $(\pi,\pi)$ spin density wave, the
$(\pi,\pi)$ $d$-charge density wave (orbital AF) and the $d-$wave pairing correlations are equally
enhanced.
However the finite size scaling does not support long range order, in complete agreement with the
results obtained by bosonization \cite{fisher}. 

\begin{figure}
  \centerline{\includegraphics[width=0.8\linewidth]{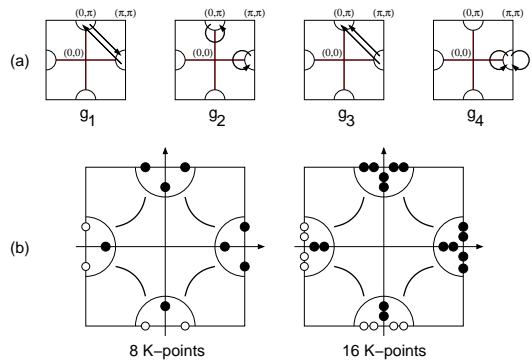}}
\vspace{-2mm}
  \caption{
    (a) The four scattering vertices of the two patch model.
    (b) The location of the $\vec{k}$-points for the 8 and 16 
    $\vec{k}$-point meshes. The curved lines denote the {\it arcs}.
    \label{fig:TwopatchSketch}}
\end{figure}

%{\em The 2D model ---} 
The $N$-patch RG analysis for the n.n.n.  2D Hubbard model revealed a 
{\em saddle point regime} \cite{honerkamp} where the FS is 
close to the SPs but sufficiently curved such that the Brillouin zone 
diagonals are not nested. 
Then the leading flow is given by the scattering 
processes between the two SP regions. The rest of the FS, which 
we will call {\em arcs}, contributes less to the flow 
%The couplings involving quasiparticles on the arcs are far less singular and get mainly  dragged along in the flow by the SP regions. At small scales, the dominant coupling between arcs and SP regions is given by Cooper pair scattering processes. 
and will be considered later. 
First let us focus exclusively on the SP regions. 
Then the flow between the SPs is the one 
described by Lederer et al.\cite{lederer} and Furukawa et al.\cite{furukawa}. There are four different types
of processes, $g_1, \dots g_4$, depicted in Fig. \ref{fig:TwopatchSketch} (a). In principle 
the scattering vertex $V_\Lambda (\vec{k}_1, \vec{k}_2,\vec{k}_3)$ 
depends on the precise location of the wave vectors $\vec{k}_i$ in the SP 
regions. As a simplification we assume that they have the same value. To a 
certain extent this can be justified by the N-patch RG. Note however that 
in the one-loop flow the couplings change smoothly from arc to SP regions.

%%%%%%%%%%%%%%%%%%%%%%%%%%%%%%%%%%%%%%%%%
%% Figure
%%%%%%%%%%%%%%%%%%%%%%%%%%%%%%%%%%%%%%%%%

\begin{figure}
  \centerline{\includegraphics*[width=0.9\linewidth]{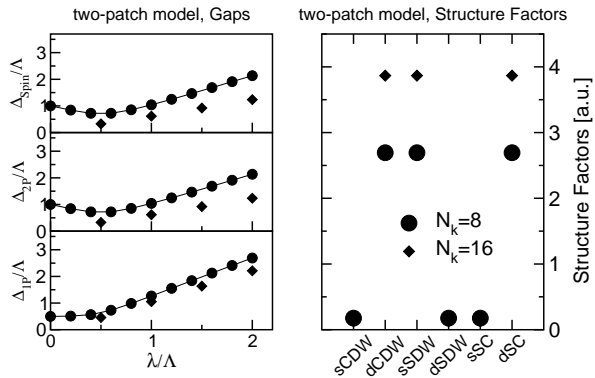}}
\vspace{-2mm}
  \caption{
    Left panel: Spin, two particle and single particle gaps in the
    two patch model as a function of the interaction parameter $\lambda$.
    Right panel: Order parameter susceptibilities for various kinds of
    order in the two patch model. No evidence for long range order is found.
    However the staggered spin, the d-wave pairing and the staggered flux
    correlations are equally enhanced with respect to the 
    noninteracting ground state.
    \label{fig:RVBTwopatch}}
\end{figure}

We implement the two patch model using the
two discretizations shown in Fig. \ref{fig:TwopatchSketch} (b). From the RG approach\cite{furukawa} 
the ratios of the couplings 
close to the critical scale are known: $g_1:g_2:g_3:g_4 = 0:1:2.2:-1$. 
Using this we calculate energy gaps and correlation functions in the same way as for the 2LHL  
case discussed above. The gap structure and structure factors are shown in Fig. \ref{fig:RVBTwopatch}.
The sSDW and dSC channels are enhanced in agreement with the divergences found in the 1-loop
RG flow \cite{furukawa} and the dCDW is also enhanced. The results show a striking analogy to the 
$d$-Mott phase of the half-filled 2LHL.
This clearly supports the idea that the initial divergence leads to the truncation of the FS
with spin and charge gaps opening near the SP. We examine next the behavior of the remaining
arcs of the non-nested FS  as the RG scale is lowered further. These arcs couple to the SP 
regions in the Cooper channel.

%%%%%%%%%%%%%%%%%%%%%%%%%%%%%%%%%%%%%%%%%
%% Figure
%%%%%%%%%%%%%%%%%%%%%%%%%%%%%%%%%%%%%%%%%

Although there is a charge gap for particle pairs $s_{\vec{p},s}$, 
$ s_{\vec{-p},s'}$ in the SP regions, quasiparticle pairs with zero total
momentum $a_{\vec{k},s}$, $a_{\vec{-k},s'}$ from the arcs can be scattered 
there as virtual excitations. The coupling can in principle be derived from 
the $N$-patch RG, its action is ($k=(i \omega, \vec{k})$)
\[
%%\begin{equation}
  S_{AS} = 
  - \frac{V_{AS} T^3}{2\Omega} \sum_{{k},{p}} g_{\vec{k}} \bar{a}_{{k},s}
  \bar{a}_{{-k},s'}  \, g_{\vec{p}} s_{-p,s'}  s_{{p},s} \, .
%%\end{equation}
\]
Here we focus on the $d_{x^2-y^2}$ channel, $g_{\vec{k}} =\cos k_x -\cos k_y $.
Formally integrating out the SP regions gives rise to an induced pairing 
interaction on the arcs
\[
%%\begin{equation}
  S_{AA}^{\mathrm{ind}} = 
  - \frac{V_{AS}^2 \chi^{d}_{S} T^3}{2\Omega} \sum_{{k},{k}'\atop s,s'} 
g_{\vec{k}}    \bar{a}_{{k},s}  \bar{a}_{{-k},s'} g_{\vec{k}'}
  a_{{-k}',s'}  a_{{k}',s} \, .
%%\end{equation}
\]
Here, $\chi ^{d}_{S}$ is the static $d$-wave pairing susceptibility per volume of the SP 
regions. $\chi ^{d}_{S}$ remains finite at all temperatures $T$, as the pair correlations 
in the $d$-Mott state decay exponentially. However the induced attraction enhances the weak
$d$-wave attraction within the arcs, and eventually, at low enough temperatures the arcs will
undergo a superconducting transition. This mechanism resembles that proposed by 
Geshkenbein et al. \cite{geshkenbein} with preformed pairs at the saddle points.
As the SP regions are insulating, the superfluid weight
will be that of the arcs alone. With a very similar reasoning for the 
$(\pi,\pi)$-processes in the spin channel we can understand how the system 
develops AF order when the arcs become more nested at sufficiently small doping.

%{\em Transition to long-range ordered states ---}
We may further ask what criterion determines whether the superfluid density
is reduced in the groundstate. The answer lies in the strength of the mutual 
reinforcement of Cooper and $(\pi,\pi)$-channels at the SPs. If the 
$(\pi,\pi)$-'almost-Cooper' processes enhance the $d$-wave pair scattering 
at the SPs more than the coupling in the Cooper channel to the arcs, we can 
expect that the system favors the reduced superfluid density. In this situation
the energy gain through the mutual reinforcement - manifest in spin 
{\em and} charge gap - is larger than the loss of superconducting condensation
energy by the partial gap to Cooper pairs in the SP regions. If the mutual 
reinforcement is not strong enough compared to other processes,
a $d$-Mott or equivalently dRVB condensate will not appear as the initial instability.
Instead a long range order such as ordinary $d$-wave superconductivity 
on the whole FS at larger doping, when the FS moves away from flat SP regions,
or AF order at small doping - sacrificing the partial spin gap - 
when the nesting over the the whole FS becomes too strong. 

%%{\em Conclusions ---} 
Summarizing we have determined a reduced fermionic Hamiltonian that contains the
essential ingredients of a $d$-Mott state. For the half-filled 
two-leg Hubbard ladder we have shown that exact diagonalization of this Hamiltonian
reproduces the known spin and charge gaps. For the 2D n.n.n. 
Hubbard model we have described how such a Hamiltonian captures the leading scattering
processes between the saddle points, and hence how a $d$-Mott condensate can truncate
the Fermi surface near the saddle points. Finally we have 
outlined how superconducting and antiferromagnetic states can arise as additional 
instabilities on the remaining parts of the Fermi surface, so that the $d$-RVB phase 
can be viewed as an unstable fixed point as proposed by Anderson \cite{anderson}.
In 2D our picture still relies on several approximations and assumptions, but we hope
that the present work may provide some guidance for further research.
\acknowledgments

We thank D.~Poilblanc and M.~Salmhofer for stimulating discussions. 
Financial support through the Swiss National Science Foundation (A.L.) 
and the German research foundation (DFG, C.H.) is acknowledged.


\vspace{-3mm}
\begin{thebibliography}{99}
\vspace{-3mm}
\bibitem{solyom}
  N. Menyh\'ard and J. S\'olyom,
  J. Low Temp. Phys. {\bf 12}, 529 (1973);
  J. Solyom, 
  Adv. Phys. {\bf 28}, 201 (1979).
\bibitem{zanchi} 
  D. Zanchi and H.J. Schulz, 
  Europhys. Lett. {\bf 44}, 235 (1998);  
  Phys. Rev. B {\bf 61}, 13609 (2000).
\bibitem{halboth} 
  C.J. Halboth and W. Metzner, 
  Phys. Rev. B {\bf 61}, 7364 (2000); 
  Phys. Rev. Lett. {\bf 85}, 5162 (2000).
\bibitem{honerkamp} 
  C. Honerkamp {\it et al.},
  Phys. Rev. B {\bf 63} 035109 (2001).
\bibitem{tflow}
  C. Honerkamp and M. Salmhofer, 
  Phys. Rev. B{\bf 64}, 184516 (2001); 
  Phys. Rev. Lett. {\bf 87}, 187004 (2001). 
\bibitem{BCS}
  J. Bardeen, L.N. Cooper and J.R. Schrieffer,
  Phys. Rev. {\bf 108}, 1175 (1957).
\bibitem{furukawa} 
  N. Furukawa, M. Salmhofer, and T.M. Rice,
  Phys. Rev. Lett {\bf 81}, 3195 (1998).
\bibitem{noack}
  R. M. Noack, S. R. White, and D. J. Scalapino,
  Phys. Rev. Lett. {\bf 73}, 882 (1994).
\bibitem{fisher}
  H.-H. Lin, L. Balents, and M. P. A. Fisher,
  Phys. Rev. B {\bf 58}, 1794 (1998).
\bibitem{dzyaloshinskii} 
  I. Dzyaloshinskii, 
  J. Phys. I (France) {\bf 6}, 119 (1996).
\bibitem{lederer}
  P. Lederer, G. Montambaux and D. Poilblanc, 
  J. Phys. {\bf 48}, 1613 (1987).
\bibitem{amlOnMethod}
  A. L\"auchli (to be published);
  A. L\"auchli, Ph.D. Thesis, ETH Z\"urich (2002).
\bibitem{geshkenbein}
  V.B. Geshkenbein, L.B. Ioffe and A.I. Larkin,
  Phys Rev. B {\bf 55}, 3173 (1996).
\bibitem{anderson}
  P.W. Anderson,
  Physica B {\bf 318}, 28 (2002).
\end{thebibliography}
\end{document}